\documentclass[twocolumn,times]{aastex63}

\newcommand{\co}{CO\,{\it J}(4$\rightarrow$3)}
\newcommand{\ci}{C\,{\sc i}\phantom{.}(1--0)}
\newcommand{\nii}{[N\,{\sc ii}]}
\newcommand{\Nii}{[N\,{\sc ii}]}
\newcommand{\bnineN}{$^{3}P_{2}$$\rightarrow$$^{3}P_{1}$}
\newcommand{\beightN}{$^{3}P_{1}$$\rightarrow$$^{3}P_{0}$}

\newcommand{\cofour}{CO\,{\it J}(4$\rightarrow$3)}

\shorttitle{[N\,{\sc ii}] at $z=2.6$}
\shortauthors{Doherty et al.}
\graphicspath{{./}{figures/}}
\usepackage{amsmath}

\begin{document}

\title{[N\,{\sc ii}] fine-structure emission at 122 and 205\,$\mu$m in a galaxy at $\text{z=2.6}$:\\ a globally dense star-forming interstellar medium}

\correspondingauthor{M.~J.~Doherty}
\email{m.doherty2@herts.ac.uk}

\author[0000-0001-8121-2432]{M.\,J.~Doherty}
\affiliation{Centre for Astrophysics Research, Department of Physics, Astronomy \& Mathematics, University of Hertfordshire, College Lane, Hatfield AL10~9AB, UK}

\author[0000-0003-4964-4635]{J.\,E.~Geach}
\affiliation{Centre for Astrophysics Research, Department of Physics, Astronomy \& Mathematics, University of Hertfordshire, College Lane, Hatfield AL10~9AB, UK}

\author[0000-0001-5118-1313]{R.\,J.~Ivison}
\affiliation{European Southern Observatory, Karl-Schwarzschild-Straße 2, D-85748 Garching, Germany}

\author[0000-0002-1318-8343]{S.~Dye}
\affiliation{School of Physics and Astronomy, University of Nottingham, University Park, Nottingham NG7~2RD, UK}




\begin{abstract}
  We present new observations with the Atacama Large
  Millimeter/sub-millimeter Array of the 122- and 205-$\mu$m fine-structure
  line emission of singly-ionised nitrogen in a strongly lensed starburst
  galaxy at $z=2.6$. The 122-/205-$\mu$m [N~{\sc ii}] line ratio is
  sensitive to electron density, $n_{\rm e}$, in the ionised interstellar
  medium, and we use this to measure
  $n_{\rm e}\approx 300\,\text{cm}^{-3}$ averaged across the galaxy. This
  is over an order of magnitude higher than the Milky Way average,
  but comparable to localised Galactic star-forming regions. Combined with
  observations of the atomic carbon (C\,{\sc i}(1--0)) and carbon monoxide (CO
  $J=4$--3) in the same system, we reveal the conditions in this intensely
  star-forming system. The majority of the molecular interstellar medium
  has been driven to high density, and the resultant conflagration of star
  formation produces a correspondingly dense ionised phase, presumably
  co-located with myriad H~{\sc ii} regions that litter the gas-rich disk.
\vspace{1cm}
\end{abstract}




\section{Introduction}

With the discovery of a population of high-redshift submillimeter-selected
galaxies (SMGs) over two decades ago
\citep{smail_deep_1997,barger_submillimetre-wavelength_1998,hughes_high-redshift_1998},
it became clear that some galaxies undergo episodes of intense star
formation in the early Universe, with rates 1000$\times$ that of the Milky
Way. Thought to be the progenitors of the most massive galaxies in the
Universe today \citep{simpson_alma_2014, toft_submillimeter_2014},
undergoing a period of rapid assembly, a central question has been how such
extreme levels of star formation are driven in these galaxies?

With the advent of sensitive, resolved submillimeter and millimeter imaging
and spectroscopy of the SMG population at high redshift \citep[see][for a
recent review]{hodge_high-redshift_2020}, a broad picture is emerging:
molecular gas reservoirs are extended on scales of several kiloparsecs
\citep{menendez-delmestre_mid-infrared_2009,
  ivison_tracing_2011,riechers_extended_2011} with a large fraction of the
molecular interstellar medium (ISM) driven to high density, driving high instantaneous
star-formation rates (SFRs) due to the increased fraction of the ISM
participating in star formation at any one time
\citep{gao_star_2004,oteo_high_2017}.

But what is the nature of star formation in these extreme systems? Is it
essentially identical to what we observe in discrete pockets in the disk of
the Milky Way, but occurring more prevalently throughout the ISM? Or is the
phenomenon of star formation in these extreme, early systems fundamentally
different --- perhaps due to systematic differences in metallicity, stellar
initial mass function, pressure, magnetic fields, and so-on? Ideally, one
would like to understand the internal conditions in more detail: the
density and phase structure, chemistry and physical distribution of the gas
in high-redshift SMGs, where we use the label `SMG' to
  broadly capture extreme star-forming galaxies.

Most progress to date has been in the exploitation of strongly
gravitationally lensed systems \citep[e.g.][]{koopmans_sloan_2006,
  swinbank_intense_2010, conley_discovery_2011, wardlow_hermes:_2013,
  dye_revealing_2015, dessauges-zavadsky_molecular_2019,rybak_full_2020}. For example,
SMM\,J2135$-$0102 ($z=2.3$), the so-called `Cosmic Eyelash'
\citep{swinbank_intense_2010} is one of the most comprehensively studied
SMGs to date, and has provided valuable insights into the conditions of the
ISM in a galaxy with an SFR roughly 100$\times$ that of the
Milky Way, close to the peak epoch of galaxy growth
\citep[e.g.][]{ivison_herschel_2010, thomson_2012, george_2014}.

\citeauthor{danielson_properties_2011} (\citeyear{danielson_properties_2011}, \citeyear{danielson_2013}) showed that the cold
molecular gas associated with star formation in the Cosmic Eyelash is
exposed to UV radiation fields up to three orders of magnitude more intense
than that of the Solar neighbourhood, with the star-forming molecular gas
having a characteristic density of $10^{4}\,\text{cm}^{-3}$, similar to
values seen in local ultraluminous infrared galaxies
\citep{davies_molecular_2003}. Recent work on another well studied lensed system, SDP\,81 \citep{rybak_full_2020}, also suggests that a large fraction of the molecular ISM has been driven to high density, with star-formation occurring throughout the galaxy, but localised in dense star-forming complexes of size $\sim$200\,pc with conditions comparable to the Orion Trapezium cluster in the Milky Way \citep[see also][]{swinbank_alma_2015}. We note, however, that the interpretation of `clumps' detected in high resolution interferometric imaging needs to be treated with caution, with \cite{ivison_giant_2020} demonstrating that previously reported star-forming clumps in the Cosmic Eyelash are in fact spurious artifacts, a result of over-cleaning of relatively low signal-to-noise data.

\begin{figure*}
    \centering
    \includegraphics[width=\linewidth]{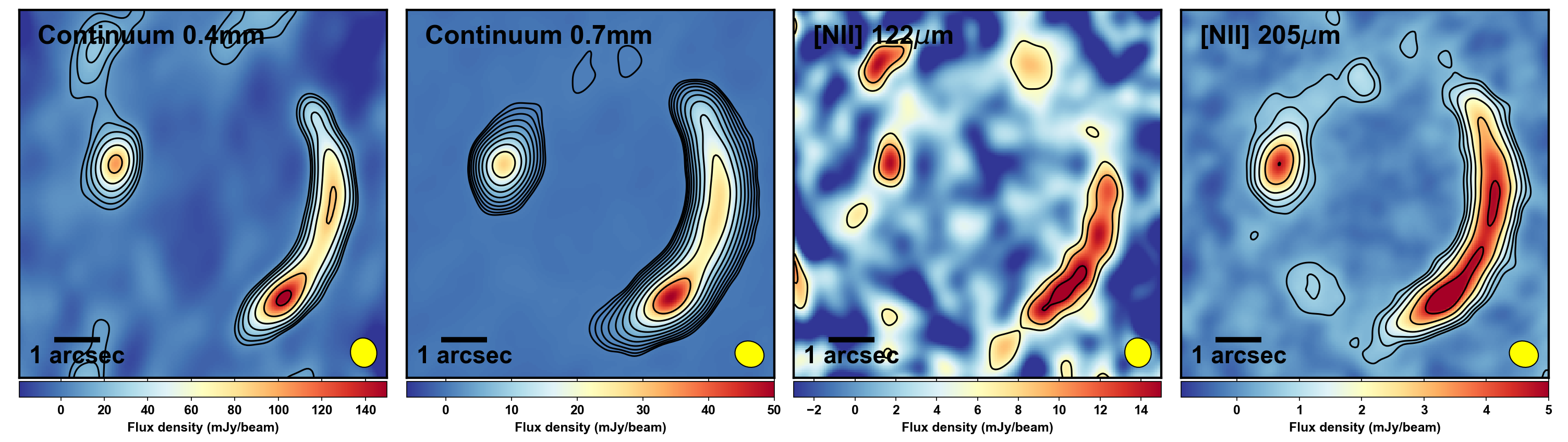}
\includegraphics[width=\linewidth]{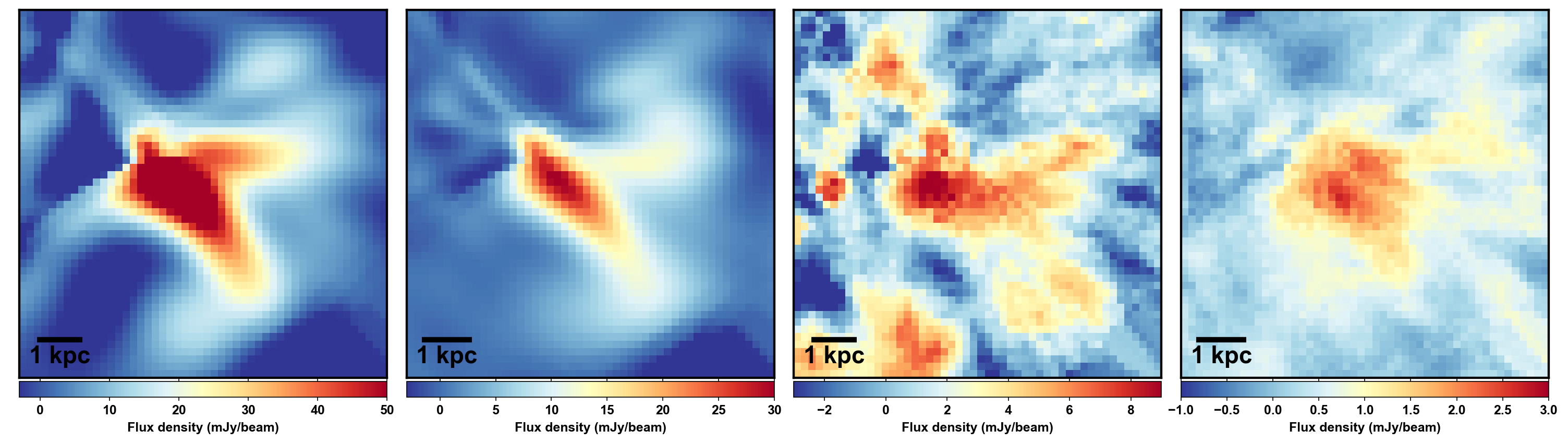}
    \caption{ALMA observations of 9io9. Top panels show the continuum
      at 0.4 and 0.7\,mm and the continuum-subtracted \Nii\ 122- and
      205-$\mu$m line emission, respectively, in the image plane, lower
     panels show the corresponding emission in the source plane. Contours in the image plane images are at logarithmically spaced (0.2\,dex) intervals of $\sigma$, starting at 3$\sigma$.}
    \label{fig:cont_and_lines}
\end{figure*}

Targets like the Cosmic Eyelash and SDP\,81 are valuable because their lens-amplified
flux offers a route to measure diagnostic tracers of the ISM and its
conditions in a way that would be impossible in the unlensed case. These
targets are, however, intrinsically rare. Here we focus on another strongly
lensed SMG: `9io9' \citep{geach_red_2015}. 9io9 is a galaxy discovered as
part of the citizen science project \textit{Spacewarps}
\citep[see][]{marshall_space_2016, more_space_2016} and independently as a
bright millimetre source by {\it Planck} \citep{harrington_early_2016},
{\it Herschel} \citep{viero_herschel_2014}, and the Atacama Cosmology
Telescope \citep{su_redshift_2017}. 9io9 lies at $z=2.6$, close to the peak
epoch of galaxy growth \citep{geach_magnified_2018}. Even after taking into
account the 15$\times$ magnification factor, 9io9 appears to have a total
infrared luminosity, $L_{\rm IR}$, exceeding $10^{13}\,\mathrm{L}_\odot$,
putting in the hyperluminous class, with an inferred SFR of around
$2000\,\mathrm{M}_{\odot}\mathrm{yr}^{-1}$, {\it modulo} the sometimes
high and often hidden AGN contribution to $L_{\rm IR}$ in such systems
\citep[e.g.][]{ivison_2019}, and also {\it modulo} evidence suggesting
that the stellar initial mass function in
high-density star-forming regions is markedly top heavy 
\citep[e.g.][cf.\ \citealt{romano_2019}]{romano_evolution_2017,
  zhang_stellar_2018, motte_2018, schneider_2018, brown_extreme_2019}.
It is therefore an excellent laboratory for studying the conditions of
intense star formation in the early Universe.

In this paper we present new observations of 9io9 with the Atacama Large
Millimeter/sub-millimeter Array (ALMA), building on the work of
\cite{geach_magnified_2018}. This study focuses on the \Nii\ 122- and
205-$\mu\text{m}$ fine-structure lines. These lines trace the cool, ionised
ISM \citep{goldsmith_herschel_2015}, and together can be used to constrain
the electron density \citep[e.g.][and references therein]{zhang_2018}. With
previous ALMA \co\ and \ci\ observations \citep{geach_magnified_2018} this
allows us to link the properties of the dense molecular and the ionised
phases of the ISM on identical scales in 9io9. In Section 2 we describe the
observations and data reduction, in Section 3 we present our analysis, in
Section 4 we interpret and discuss the main results of the analysis and
present our conclusions in Section 5. Throughout we assume a `Planck 2015'
cosmology where $H_{0} = 68~\text{km}~\text{s}^{-1}\,\text{Mpc}^{-1}$ and
$\Omega_{\text{m}} = 0.31$ \citep{planck_collaboration_planck_2016}.

\section{Observations}
\label{sec:Observations} 

9io9 ($02^{\rm h}09^{\rm m}41\fs3$, $00\degr15\arcmin58\farcs5$,
$z=2.5543$) was observed with the ALMA 12-m array in project 2017.1.00814.S
in Bands 4, 8 and 9. The details of the Band 4 observations are presented
in \cite{geach_magnified_2018}, and here we present the new Band 8 and 9
data. The Band 8 observations were conducted on 2018 August 26 and 2018
September 07 in two 40-min execution blocks with a representative frequency
of approximately 411.5~GHz. The antenna configuration was C43--4 (with 47
antennas), with a maximum baseline of 783\,m. The central frequencies of
the four spectral windows were 397.466, 399.404, 409.481 and 411.481\,GHz,
designed to cover the the redshifted \nii\ \beightN{} fine-structure line
($\nu_{\text{rest}} = 1461.131$\,GHz) and continuum emission. Calibrators
included J0217+0144, J0423$-$0120 and J2253+1608. In two executions the
total on-source integration time was 42\,min, with an average precipitable
water vapor column of 0.73\,mm, resulting in an r.m.s.\ noise of
3\,mJy~beam$^{-1}$ in a 15.6-MHz channel.

The Band 9 observations were conducted on 2018 August 19 in a single 70-min
execution block. The antenna configuration was C43--3 (with 46 antennas),
with a maximum baseline of 484\,m. The representative frequency was
approximately 692.3\,GHz with eight spectral windows at 670.311, 672.249,
674.249, 676.202, 686.248, 688.201, 690.200 and 692.138\,GHz. The Band 9
observations were designed to contain the redshifted \nii\ \bnineN{}
fine-structure line ($\nu_{\text{rest}} = 2459.380$\,GHz) and continuum
emission. Calibrators included J0217+0144 and J0522$-$3627. The total time
on-source was 31\,min, with an average precipitable water vapor column of
0.45\,mm, resulting in an r.m.s.\ noise of 10~mJy~beam$^{-1}$ in a 15.6-MHz
channel.

For the Band 8 data we used the pipeline-restored calibrated measurement
set. For the Band 9 observations, due to unstable phases in two antennas
(DA64 and DA24) we re-calibrated manually, flagging both dishes within the
pipeline script. We imaged and cleaned the data using the {\sc casa}
(v.5.1.0-74.el7) task, {\tt tclean}. As in \cite{geach_magnified_2018}, we
employ multi-scale cleaning (at scales of $0\arcsec, 0.5\arcsec,
$and$, 1.25\arcsec$). First we produce dirty cubes to establish the r.m.s.
(1$\sigma$) noise per channel, and then clean down to a threshold of
3$\sigma$. With natural weighting, the synthesised beams were
$0.43\arcsec \times 0.35\arcsec$ (FWHM, position angle $33\degr$) and
$0.32\arcsec \times 0.28\arcsec$ (position angle $74\degr$) in Bands 8 and
9, respectively. In order to produce data cubes that have closely matched
resolution across Band 4, 8 and 9, we also produce {\it uv}-tapered maps
with a scale of 0.6$''$ noting that the {\it uv} plane is well sampled across all bands such that we expect negligible losses to extended emission on the scale of the lens (LAS $\lesssim$3--4$''$). In Fig.~\ref{fig:cont_and_lines} we present the
image-plane maps of the line-free Band 8 (400\,GHz) and Band 9 (680\,GHz)
continuum emission and the line-averaged \nii\ \beightN\ and \bnineN\
continuum-subtracted emission.

\section{Analysis}

\subsection{Lens model}

\begin{figure}
    \centering
    \includegraphics[width=\columnwidth]{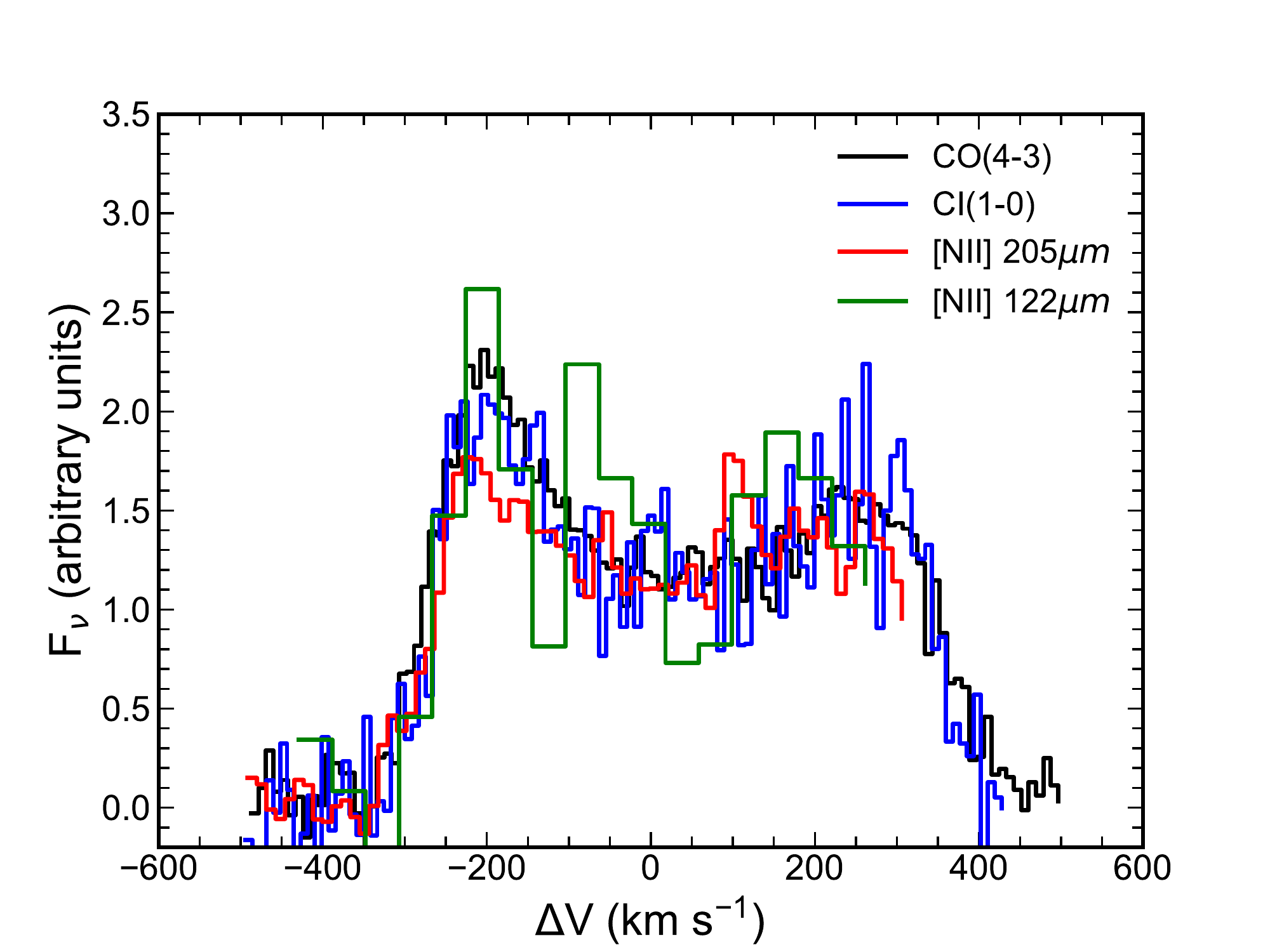}
    \caption{\Nii\ doublet, \ci\ and \co, scaled such that each line has a common mean (on an arbitrary scale). All lines share a very similar double horned profile, indicating that the molecular and ionized phases are broadly tracing the same structures.}
    \label{fig:scaled_emission}
\end{figure}

We adopt the same lens model as \cite{geach_magnified_2018}. Briefly, the
lens model includes the gravitational potential of both the primary lensing
galaxy ($z\approx0.2$) and its smaller northern companion (assumed to be at
the same redshift). The model uses the semi-linear inversion method of
\cite{warren_semilinear_2003} to reconstruct the source plane surface brightness that best matches the
observed Einstein ring for a given lens model. This process is iterated,
varying the lens model and creating a source reconstruction at each
iteration, until the global best fit lens model is found \citep{geach_magnified_2018}. The best fit model was used to produce source-plane
cubes. Each slice was reconstructed with 50 realisations of a randomised Voronoi source plane grid \citep[see][for further details]{dye_modelling_2018} and the mean taken, weighted by the log of the likelihood of each realisation. In Fig.~\ref{fig:cont_and_lines} we show the equivalent continuum
and line maps in the source plane. The resulting source plane
reconstruction has an average beam size of $280\,\text{mas}$ (FWHM)
corresponding to a physical scale of 2.3\,kpc in the source plane.
Fig.~\ref{fig:cont_and_lines} shows the source-plane images of the
line-free Band 8 (400\,GHz) and Band 9 (680\,GHz) continuum emission and
the line-averaged \nii\ \beightN\ and \bnineN\ continuum-subtracted
emission. As can be seen in Figure~\ref{fig:cont_and_lines}, the \Nii{} $122\mu$m line measured in Band 9 appears to have an additional clump of emission to the NE not clearly visible (however detected) in the $205\mu$m map. This feature is multiply imaged in the image plane and also detected in the Band 8 data, so although at low significance unlikely to be spurious. However due to a lack of a continuum detection and the low significance of the detection, the strength of the $122\mu$m line in this region is unreliable. There are several other low significance features that do correlate with emission in other bands along the ring, but due to the relatively low signal-to-noise of the $122\mu$m line we do not attempt to interpret the resolved properties of the line itself, or the line ratio, concentrating our analysis on the integrated properties. Of course, with the $122\mu$m line strength now constrained, future observations could be obtained to address the resolved properties at higher signal-to-noise. Fig.~\ref{fig:scaled_emission} shows the \Nii\ doublet as well as
the \ci\ and \co\ lines, scaled for comparison. In the following, all
analysis is performed in the source plane. The errors quoted for the
intrinsic source properties do not include any systematic uncertainty
introduced by the prescribed parametric lens model \citep[see
e.g.][]{schneider_mass-sheet_2013}.

\subsection{Line and continuum measurements}

As we have resolved the global thermal dust continuum emission from 9io9 at
high signal-to-noise in each of ALMA Bands 4, 8 and 9, with Band 9
probing close to the redshifted peak of the thermal emission, we can
estimate $L_{\rm IR}$ (rest-frame 8--1000\,$\mu$m) for the source, fully
taking into account the effects of differential lensing since $L_{\rm IR}$ can be determined in the source plane. We fit the observed emission with
a single-temperature modified blackbody spectrum, allowing the dust
emissivity ($\beta$), dust temperature ($T_{\rm d}$) and normalisation to
be free parameters. Summing the luminosity within the region bound by the
$\geq$3$\sigma$ contour (in Band 8) we measure a total
$L_{\rm IR}=(1.1\pm 0.2)\times 10^{13}\,\mathrm{L}_\odot$. Thus, 9io9 is a
hyperluminous infrared-luminous galaxy (HyLIRG) at $z\approx2.6$. For
reference, the total Band 8 and Band 9 source plane continuum flux
densities are 12\,mJy and 43\,mJy at observed frequencies of 400 and
680\,GHz, respectively.

We evaluate line luminosities using the standard relation \cite{solomon_molecular_1997}
\begin{equation}
\frac{L}{\mathrm{L}_\odot} = \left(\frac{1.04 \times 10^{-3}\nu_{\text{obs}}}{\text{GHz}}\right) \left(\frac{D_{\text{L}}}{\text{Mpc}}\right)^{2}\left(\frac{S\Delta V}{\text{Jy~km~s}^{-1}}\right)
\label{eq:line_lum}
\end{equation}

\noindent
where $D_{\text{L}}$ is the luminosity distance, $\nu_{\text{obs}}$ is the
observed frequency and $S\Delta V$ is the velocity-integrated line flux.
Subtracting the continuum model from each pixel based on a linear fit to line free regions of the lines corresponding data cube allowing the gradient and normalisation vary as free parameters, the integrated flux density (measured within the same $3\sigma$ region as the continuum) of the \Nii\ 205-$\mu$m line is $\mathbf{S}\Delta V = (2.37 \pm 0.06)~\text{Jy~km}~\text{s}^{-1}$,
corresponding to $L_{205}=(4.7 \pm 0.1) \times 10^{8} \,\mathrm{L}_\odot$.
For the \Nii\ 122-$\mu$m line, the integrated flux density is
S$\Delta V = (7.6 \pm 0.9)~\text{Jy~km}~\text{s}^{-1}$, corresponding to
$L_{122}=(2.6 \pm 0.3) \times 10^{9} \,\mathrm{L}_\odot$, integrating between $-500~\text{km}~\text{s}^{-1}$~and $300~\text{km}~\text{s}^{-1}$ due to the \Nii{} lines being incomplete. By comparing to the \cofour{} and \ci{} lines we estimate we are missing $\approx 10\%$ of the total line flux for both lines, however this should not alter the resultant \nii{} ratio as both lines are equally affected. Uncertainties are
determined by adding Gaussian noise to each channel, with a scale
determined from off-source regions of the datacube, taking the standard
deviation of the pixel-to-pixel channel noise and scaling by the solid
angle subtended by the 3$\sigma$ mask used to define the total emission.
Repeating this 1000 times and integrating the lines for each realisation
allows us to estimate the uncertainties.

\subsection{Electron density}

With an ionisation potential of 14.5\,eV, N$^+$ originates exclusively from
the ionised ISM. N$^+$ thus traces H\,{\sc ii} regions and is a good tracer
of massive star formation, with its emission being directly related to the
ionising photon rate and thus the number of massive stars
\citep{zhao_[nii]_2016, herrera-camus_ionized_2016}. At lower densities
($0.01 \leq n_{\text{e}} \leq 0.1~\text{cm}^{-3}$), N$^+$ is also a coolant
of the warm ionised medium (WIM).

N$^+$ has three fine-structure levels: $^3P_{0,1,2}$. These are simply
referred to as 0, 1, 2 in the following. If electron collisions are the
primary excitation mechanism, we can write the collision rates as
$C_{\text{ij}} = R_{\text{ij}}n_{\text{e}}$ where $R_{\text{ij}}$ is the
collision rate coefficient and $n_{\text{e}}$ is the electron density.
Following \cite{goldsmith_herschel_2015}, the rate equations can be written

\begin{eqnarray}
\begin{aligned}
    -(A_{21} + C_{21} + C_{20})n_{2} + C_{12}n_{1} + C_{02}n_{0} = 0  \\
    (A_{21} + C_{21})n_{2} -(A_{10} + C_{10} + C_{12})n_{1} + C_{01}n_{0} = 0   \\
    C_{20}n_{2} + (A_{10} + C_{10})n_{1} - (C_{02} + C_{01})n_{0} = 0 \\
    \label{eq:rate_equations}
\end{aligned}
\end{eqnarray}
\noindent
where $n_{0}$, $n_{1}$, and $n_{2}$ are the populations of each
level, such that $n_{0} + n_{1} + n_{2} = n(\text{N}^{+})$, and
$A_{\text {ij}}$ are the Einstein coefficients. The 122- and 205-$\mu$m
emission lines correspond to energy transitions $E_{21}$ and $E_{10}$,
respectively. For optically thin emission these are related to the line
intensity of the $\text{i}\rightarrow\text{j}$ transition as
\begin{equation}
  I_{\text{ij}} = \frac{A_{\text{ij}}E_{\text{ij}}N_{\text{i}}}{4\pi},
\end{equation}
where $N_{\text i}$ is the column density of the upper level.
The ratio of the line intensities then becomes
\begin{equation}
   R= \frac{I_{21}}{I_{10}} = \frac{A_{21}E_{21}n_{2}}{A_{10}E_{10}n_{1}}, 
    \label{eq:intensity_rat2}
\end{equation}
assuming the ratio of column densities equals the ratio of volume
densities. The rate equations (\ref{eq:rate_equations}) yield

\begin{equation}
    \frac{n_{2}}{n_{1}} = \frac{C_{12}(C_{01} +C_{02})  + C_{02}(A_{10} + C_{10})}{(A_{21} + C_{21} + C_{20})(A_{10} + C_{10}) + C_{20}C_{12}}.
\end{equation}

\begin{figure}
    \centering
    \includegraphics[width=\columnwidth]{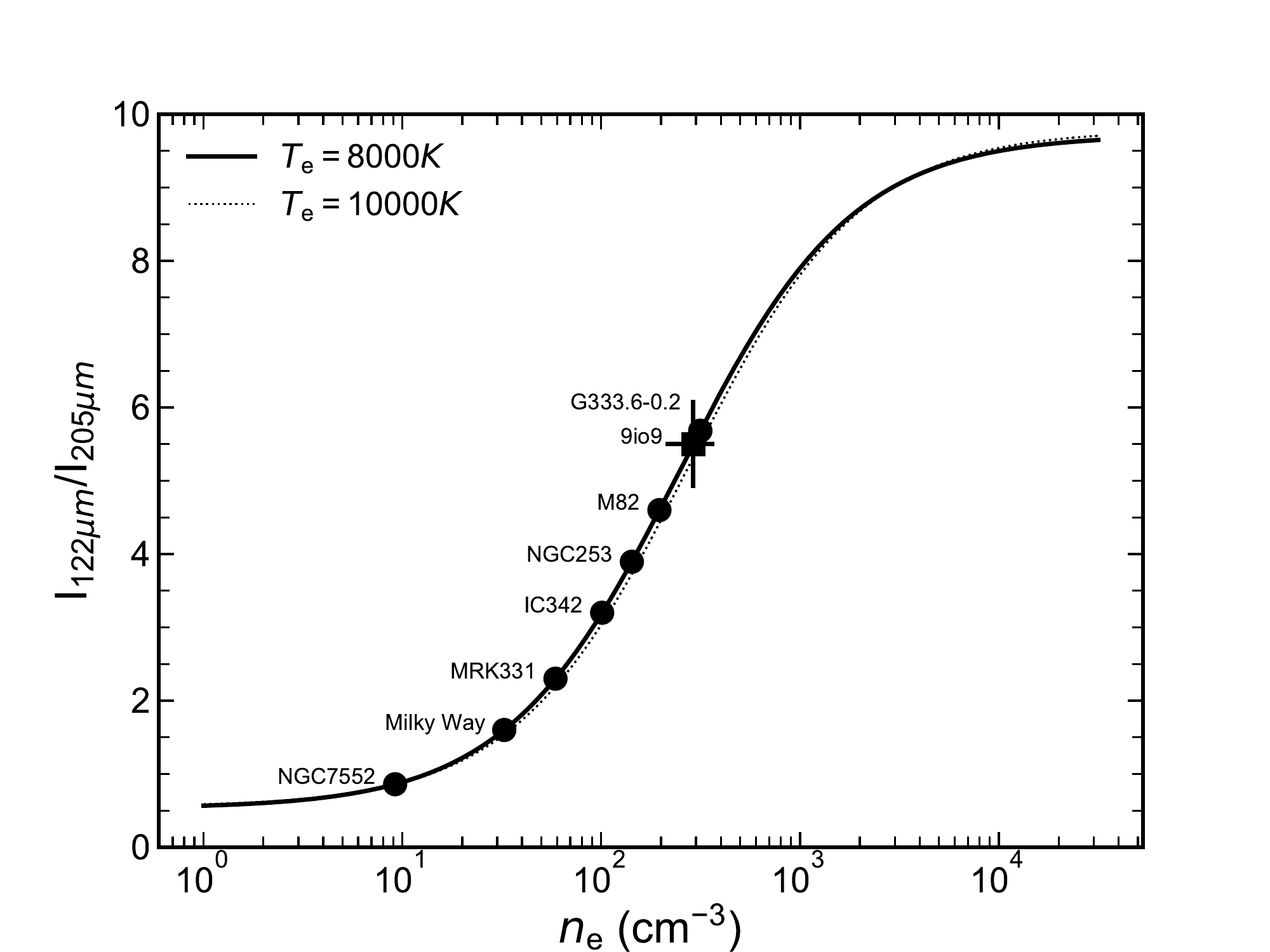}
    \caption{Derived electron density versus \Nii\ line ratio, including literature values found for a variety of Galactic and extragalactic sources. Note the insensitivity of $n_{\text{e}}$ to the assumed electron temperature.}
    \label{fig:ne_relation}
\end{figure}

Using $C_{\text{ij}} = R_{\text{ij}}n_{\text{e}}$, we derive a relation for
the electron density which can be written in the form \begin{equation}
    n_{\text e} = \frac{d}{c}\frac{R - R_{\text{min}}}{R_{\text{max}} -R},
\end{equation}

where 
\begin{eqnarray}
\begin{aligned}
    a = {R_{12}R_{01} + R_{02}R_{10} + R_{02}R_{12}}  \\ 
    b = R_{02}A_{10} \\
    c = R_{02}R_{21} + R_{01}R_{21} + R_{01}R_{12} \\
    d = A_{21}(R_{02} + R_{01})\\
    R_{\text {min}} = \frac{A_{21}E_{21}}{A_{10}E_{10}}\frac{b}{d} \\
    R_{\text{max}} = \frac{A_{21}E_{21}}{A_{10}E_{10}}\frac{a}{c}. \\ 
\end{aligned}
\end{eqnarray} Using collision rates from
\cite{tayal_electron_2011} for a kinetic temperature,
$T_{\text e} = 8000$\,K, and assuming that collision rates are independent
of temperature\footnote{Although the relation is not completely temperature
  independent, the derived electron density does not vary by more than 10\%
  for a 2000\,K increase in kinetic temperature, as can be seen in
  Fig.~\ref{fig:ne_relation}.} yields an expression for the electron
density based on the line ratios: \begin{equation} \label{eq:ne247}
    n_{\text{e}}=247\left(\frac{R - 0.52}{9.73-R}\right)\text{cm}^{-3}.
\end{equation}

Using our measurements of the line intensities, we find an electron
density of $n_{\text{e}} = 290^{+90}_{-70}\text{cm}^{-3}$. 
This elevated electron density could explain the discrepancy found between
the \Nii\ $205\mu$m and $L_{\rm IR}$-derived SFRs by
\citet{harrington_red_2019}, since in that work a low electron density ---
$n_{\rm e}\approx 30~\text{cm}^{-3}$ --- was assumed.

There may be a potential bias in our measurement if the local radiation
field is so intense that optical pumping of \Nii\ provides a significant
contribution to the excitation \citep{flannery_ultraviolet_1979}. We can
test if this is likely to be important by comparing the pumping collision
rate with the inferred collision rate from our data. The pump rate can be
expressed as $R_{\text{p}} = 1835 \nu U_\nu$, where $\nu U_\nu$ is a
measure of the strength of the radiation field in erg~cm$^{-3}$. In the
Solar neighbourhood, $\nu U_\nu\approx 7 \times 10^{-14}$~erg~cm$^{-3}$ at
$\nu\approx3 \times 10^{15}$\,Hz corresponding to the
$^{3}{D}_{1}\rightarrow^{3}{P}_{0}$ transition \citep{draine_physics_2011}.
The interstellar UV radiation field in galaxies like 9io9 is not well
constrained, however \citet{danielson_properties_2011} found for the Cosmic
Eyelash that the cold gas was exposed to a UV radiation field that is
roughly 1000$\times$ that of the local Galactic ISM. If we assume similar
conditions for 9io9, then we obtain a pump rate of
$R_{\text{p}} \approx 10^{-7}\,\text{s}^{-1}$, or approximately 2 per cent
of the collision rate. Thus, even with rather extreme local radiation
fields --- perhaps only relevant for the gas in the immediate vicinity of O
and B stars --- pumping has a negligible impact on the \Nii\ excitation.

\begin{figure}
    \centering
    \includegraphics[width=\columnwidth]{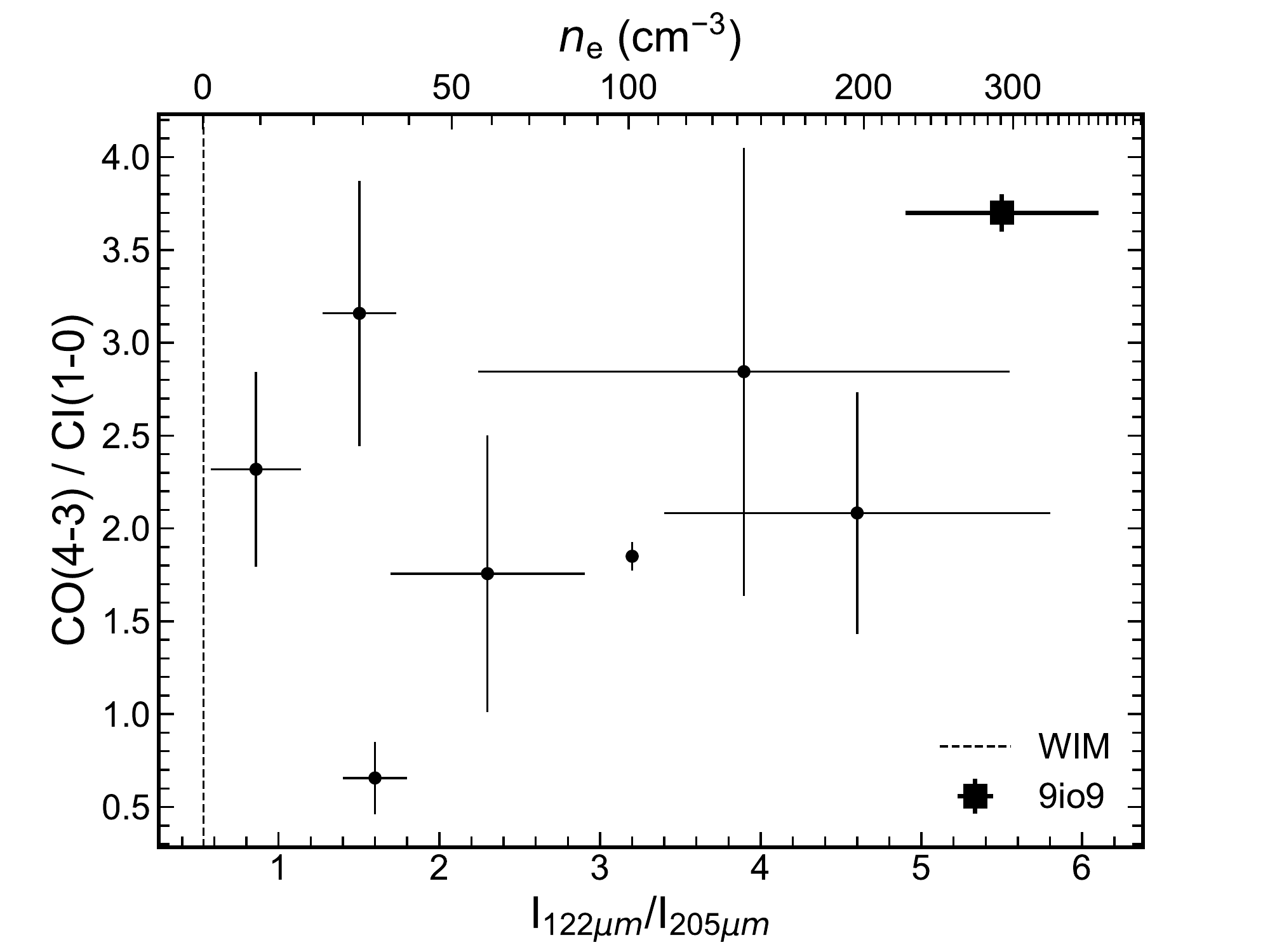}
    \caption{\Nii\ line ratio versus \co/\ci{} (a dense to total molecular gas mass tracer)
      for a range of sources
      \citep{wright_preliminary_1991,petuchowski_n_1994,zhang_co_2001,oberst_detection_2006,rigopoulou_herschel-spire_2013,rosenberg_radiative_2014,rosenberg_theherschelcomprehensive_2015,zhao_[nii]_2016}.}
    \label{fig:ne_co_ci}
\end{figure}

\section{Interpretation and discussion}

Recent studies of electron densities in the ISM of star-forming galaxies at
high redshift using optical/near-infrared (NIR) tracers of $n_{\text{e}}$
--- such as O\,{\sc ii} and S\,{\sc ii} --- have also found values
significantly higher than those typically observed in the local Universe,
with $n_{\rm e}\gtrsim100$\,cm$^{-3}$ not uncommon in samples of
(H$\alpha$-selected) star-forming galaxies with
$\text{SFR}\sim 1$--$100\,\text{s}~\text{M}_{\odot}\,\text{yr}^{-1}$
\citep{masters_physical_2014,shirazi_stars_2014}.
\citet{kaasinen_cosmos_2017} connected these elevated electron densities to
the high SFRs observed in galaxies at \smash{$z\sim1$--$2$}, pointing out
that the average electron densities in their high-redshift and local
samples are comparable, when controlled for SFR.
\citet{shimakawa_correlation_2015} and \citet{jiang_correlation_2019} note a
correlation between the surface density of star formation
(\smash{$\Sigma_{\text{SFR}}$}) and $n_{\text{e}}$, providing further
evidence of the close relationship between star formation and ionised ISM
density. This is not surprising, since if star formation depends on the
density distribution of the cold ISM, and $\Sigma_{\text{SFR}}$ is enhanced
when a large fraction of the molecular ISM is driven to high density, then
that will likely be reflected in the resultant ionised phase density in the
vicinity of those star-forming regions.

The integrated profiles of both \Nii\ lines resemble the integrated \co\
and \ci\ line profiles (recall Fig.~2), indicating that the ionised and
molecular tracers originate from the same material and local environments.
\citet{geach_magnified_2018} modelled the \co\ emission with a rotating
disk motivated by the dynamics of circumnuclear gas seen in local ULIRGs
\citep{downes_rotating_1998}, and this provided an excellent fit to the
data, with properties similar to another stongly lensed system, SDP\,81
\citep[e.g.][]{alma_2015, swinbank_alma_2015, dye_2015,
  hatsukade_high-resolution_2015,rybak_full_2020}. It is reasonable to assume that the
\Nii\ emission is tracing the ionised component of the ISM across the same
disk, and is broadly co-spatial with the star-forming molecular gas. The
measured electron density of
$n_{\text{e}} = 290^{+90}_{-70}\,\text{cm}^{-3}$ is significantly higher
than that expected for the WIM, which is comprised of material at densities
3--4 orders of magnitude lower
\citep{gaensler_vertical_2008,weisberg_arecibo_2008}.

The electron density in 9io9 is high, but not extreme when we consider
individual local star-forming environments. For example, the Galactic
H\,{\sc ii} region, G333.6$-$0.2, has an electron density,
$\approx300$\,cm$^{-3}$, also determined via the \Nii\ ratio
\citep{colgan_detection_1993}. This is consistent with the conditions in
9io9, with the key difference that we are measuring the characteristic
electron density on scales of several kiloparsecs, rather than for an
individual star-forming complex. Thus, a possible scenario is that star
formation in 9io9 is proceeding in environments that resemble `normal'
star-forming regions like G333.6$-$0.2 but with the key distinction that,
while G333.6$-$0.2 represents a tiny fraction of the total Milky Way ISM by
mass and volume, in 9io9 the {\it majority} of the ISM may be in this
state.

There is support for this `globally dense' picture in the molecular phase
tracers. \cite{papadopoulos_molecular_2012} argue that the \co{}/\ci\ ratio
is an excellent empirical tracer of the ratio of dense, actively
star-forming molecular gas to the total molecular reservoir.
\cite{geach_magnified_2018} show that the observed ratio in 9io9 is
consistent with over half of the molecular ISM having been driven to high
density. It follows that the H\,{\sc ii} regions produced by massive star
formation should have correspondingly elevated electron densities. Hints of
this link can be seen in Fig.~\ref{fig:ne_co_ci}, where we compare the
\co{}/\ci\ ratio to the \Nii\ line ratio for 9io9 and a sample of sources
from the literature spanning a range of densities. Generally, there is a
weak trend that galaxies or regions with high electron densities --- as
traced by the \Nii\ line ratio --- have a correspondingly high dense
molecular gas fraction --- as traced by the \co{}/\ci\ ratio.

Given 9io9’s clear disk-like structure, most evident in the shape of the
line profiles and confirmed by the excellent fits achieved for disk- or
ring-like kinematic models \citep{geach_magnified_2018}, we can consider a
detailed picture of star formation in this galaxy. A plausible scenario is
that the disk comprises an ensemble of dense clumps containing a large
fraction of the total ISM, within which star formation occurs. This is not
a novel concept, of course. Early work with {\it Hubble
  Space Telescope} suggested that a large fraction of star formation in
high-redshift Lyman-break galaxies may occur in large blue clumps, on
scales of up to 1\,kpc \citep[e.g.][]{cowie_faintest_1995,
  elmegreen_galaxy_2005}. More recently, resolved imaging of strongly
lensed dusty star-forming systems suggests the presence of
distinct regions of embedded high-density star formation on scales of
$\sim$100\,pc, with luminosity densities comparable to the cores of local
giant molecular clouds \citep[e.g.][though see 
\citealt{ivison_giant_2020}]{swinbank_intense_2010, swinbank_alma_2015,
  hatsukade_high-resolution_2015}. This led to a `giant clump' model of
star formation in the most vigorously star-forming galaxies at high
redshift, where the 100\,pc-scale clumps thought to be present in objects
like the Cosmic Eyelash resemble scaled-up versions of the dense,
1\,pc-scale cores within local giant molecular clouds.

The physical argument put forward to explain the formation of such
structures is through disk instabilities, which conveniently circumvents
the requirement for mergers or interactions to drive gas to high densities,
although interactions are known to be common --- perhaps even ubiquitous
--- amongst SMGs \citep[e.g.][]{engel_2010}. At high gas fractions and
surface densities, disks will be Toomre-unstable and undergo local collapse
\citep{toomre_gravitational_1964, noguchi_early_1999,
  dekel_formation_2009}. The Jeans length scale on which this fragmentation
occurs for the typical gas densities and velocity dispersions {\it
  inferred} in objects like the Cosmic Eyelash, SDP\,81 and 9io9 is broadly
consistent with the $\approx$100\,pc clump scales described above
\citep[e.g.][]{swinbank_alma_2015, hatsukade_high-resolution_2015}\footnote{Note that in the case of the Cosmic Eyelash \cite{ivison_giant_2020} have shown these clumps to be spurious ---
the result of over-cleaning a low-signal-to-noise interferometric image.}. The
current source plane resolution in 9io9 is around 300\,pc and so we cannot
yet address this question; however, the brightness of the target
(approaching 1\,Jy) makes it a prime candidate for pushing to very long
baselines to resolve the disk sub-structure, allowing us to link the
physical conditions of the ISM explored here on global scales to the
spatial distribution of the gas down to 10s of parsecs.

\section{Conclusions}

We have reported new ALMA Band 8 and 9 observations of a strongly lensed
HyLIRG at $z = 2.6$, targeting the \nii\ fine-structure line emission and
thermal continuum in the rest-frame far-infrared waveband. Our main
findings are as follows:

\begin{itemize}
\item We report detections of both the 122- and 205-$\mu$m \nii\ emission
  lines, which trace the ionised ISM. The \nii\ lines match the
  double-horned line profiles seen in \co\ and \ci, reported by
  \cite{geach_magnified_2018} and well-modelled by a rotating
  disk. This implies that the ionised gas is broadly co-located with the
  molecular material on scales of several kiloparsecs.
\item We use the 122/205$\mu$m line ratio to estimate the average electron
  density in the ISM, finding $n_{\rm e}\approx 300\,\mathrm{cm}^{-3}$, an
  order of magnitude above that of the (average) Milky Way, comparable
  with measurements of the electron density in discrete Galactic
  star-forming  environments.
\item We demonstrate a tentative (but expected) trend between the ratio of
  dense molecular gas and the total molecular gas reservoir, as traced by \co\
  and \ci{}, and the \nii\ line ratio. If the former is a tracer of the
  dense molecular gas fraction as \cite{papadopoulos_molecular_2012} argue,
  then the correlation with the \nii\ ratio and therefore ionised gas
  density reveals a picture of `globally dense' ISM, where a significant
  fraction of the molecular component has been driven to high density ---
  possibly through violent disk instabilities, with or without galaxy
  interactions --- with myriad individual
  H\,{\sc ii} regions dominating the observed \nii\ emission.
\end{itemize}

9io9 is a remarkably extreme system, fortuitously lensed to provide us a
glimpse of its inner workings. Our findings support a picture where the
nature of star formation in this galaxy might not necessarily differ from
the conditions of star formation in our own Milky Way. The key difference
is that a far higher fraction of the ISM is currently participating in that
star formation.

\section*{Acknowledgements}

M.J.D.\ and J.E.G.\ are supported by the Royal Society. S.D.\ is supported by an STFC
Rutherford Fellowship. This paper makes use of the following ALMA data:
ADS/JAO.ALMA\#2017.1.00814.S. ALMA is a partnership of ESO (representing its
member states), NSF (USA) and NINS (Japan), together with NRC (Canada),
MOST and ASIAA (Taiwan), and KASI (Republic of Korea), in cooperation with
the Republic of Chile. The Joint ALMA Observatory is operated by ESO,
AUI/NRAO and NAOJ. Funded by the Deutsche Forschungsgemeinschaft (DFG, German Research
Foundation) under Germany's Excellence Strategy --- EXC-2094 ---
390783311. This research has made use of the University of
Hertfordshire high-performance computing facility
(\url{http://stri-cluster.herts.ac.uk}).

\bibliography{references}{}
\bibliographystyle{aasjournal}

\end{document}